\documentclass[hidelinks,conference,letterpaper]{IEEEtran}

\usepackage{bm}
\usepackage{amsmath}
\usepackage{extarrows}
\usepackage{amssymb}
\usepackage{graphicx}
\usepackage{color, xcolor}
\usepackage{enumerate}
\usepackage{bookmark}
\graphicspath{{figure}}
\usepackage{setspace}
\usepackage{multirow}
\usepackage{amsmath}
\usepackage{subfigure}
\usepackage{float}
\usepackage{booktabs}
\usepackage{algorithm}
\usepackage{algorithmic}
\usepackage{cite}
\usepackage{framed} 
\usepackage{scalerel}

\newcommand{\mr}{\mathrm}

\newcommand{\BE}{\begin{equation}}
\newcommand{\EE}{\end{equation}}
\newcommand{\BS}{\begin{subequations}}
\newcommand{\ES}{\end{subequations}}
\renewcommand{\bf}{\bm}
\newcommand{\tabincell}[2]{\begin{tabular}{@{}#1@{}}#2\end{tabular}}
\newtheorem{theorem}{Theorem}
\newtheorem{proposition}{Proposition}

\newtheorem{definition}{Definition}

\newtheorem{lemma}{Lemma}

\begin{document}

\title{Memory Approximate Message Passing} 
\author{Lei~Liu,   Shunqi~Huang and  Brian~M.~Kurkoski\\
  School of Information Science, Japan Institute of Science and Technology, Japan}

% \author{\IEEEauthorblockN{Lei~Liu, \emph{Member, IEEE},  Shunqi~Huang, Yiyao Cheng, and\\ Brian~M.~Kurkoski, \emph{Member, IEEE}}
% \thanks{Lei~Liu, Shunqi Huang and Brian~M.~Kurkoski are with the School of Information Science, Japan Institute of Science and Technology (JAIST), Nomi 923-1292, Japan (e-mail: \{leiliu, s1910413, kurkoski\}@jaist.ac.jp).} 
%  \thanks{Yiyao~Cheng is with the Department of Electrical Engineering, City University of Hong Kong, Hong Kong, SAR, China (e-mail:  yiycheng2-c@my.cityu.edu.hk).}}

\maketitle

\begin{abstract}
Approximate message passing (AMP) is a low-cost iterative parameter-estimation technique for certain high-dimensional linear systems with non-Gaussian distributions. However, AMP only applies to independent identically distributed (IID) transform matrices, but may become unreliable for other matrix ensembles, especially for ill-conditioned ones. To handle this difficulty, orthogonal/vector AMP (OAMP/VAMP) was proposed for general right-unitarily-invariant matrices. However, the Bayes-optimal OAMP/VAMP requires high-complexity linear minimum mean square error estimator. To solve the disadvantages of AMP and OAMP/VAMP, this paper proposes a memory AMP (MAMP), in which a long-memory matched filter is proposed for interference suppression. The complexity of MAMP is comparable to AMP.  The asymptotic Gaussianity of estimation errors in MAMP is guaranteed by the orthogonality principle. A state evolution is derived to asymptotically characterize the performance of MAMP. Based on the state evolution, the relaxation parameters and damping vector in MAMP are optimized. For all right-unitarily-invariant matrices, the optimized MAMP converges to OAMP/VAMP, and thus is Bayes-optimal if it has a unique fixed point. Finally, simulations are provided to verify the validity and accuracy of the theoretical results.
\end{abstract}

 \footnotetext{
 A full version of this paper is accessible at
\href{https://arxiv.org/pdf/2012.10861.pdf}{arXiv} (see \cite{MAMParxiv}). The source code of this work is publicly available at 
\href{https://sites.google.com/site/leihomepage/research}{sites.google.com/site/leihomepage/researc}.\par 

L. Liu was supported in part by the Japan Society for the Promotion of Science (JSPS) Kakenhi under Grant JP 21K14156, and in part by JSPS Kakenhi Grant JP 19H02137.
 
}    

\section{Introduction}

Consider the problem of signal reconstruction for a noisy linear system:
\BE \label{Eqn:linear_system}
\bf{y}=\bf{Ax}+\bf{n} 
\EE
where $\bm{y}\!\in\!\mathbb{C}^{M\!\times\!1}$ is a vector of observations, $\bf{A}\!\in\!\mathbb{C}^{M\!\times\! N}$ is a transform matrix, $\bf{x}$ is a vector to be estimated and $\bm{n}\!\sim\!\mathcal{CN}(\mathbf{0},\sigma^2\bm{I}_M)$ is a vector of Gaussian additive noise samples. The entries of $\bf{x}$ are independent and identically distributed (IID) with zero mean and unit variance, i.e., $x_i\sim P_x$. In this paper, we consider a large system with $M,N\to\infty$ and a fixed $\delta=M/N$ (compressed ratio). In the special case when $\bf{x}$ is Gaussian, the optimal solution can be obtained using standard linear minimum mean square error (MMSE) methods. Otherwise, the problem is in general NP hard \cite{Micciancio2001,verdu1984_1}.

\subsection{Background}
Approximate message passing (AMP) has attracted extensive research interest for this problem \cite{Bayati2011, Donoho2009}. AMP adopts a low-complexity matched filter (MF),  so its complexity is as low as ${\cal O}(MN)$ per iteration. Remarkably, the asymptotic performance of AMP can be described by a scalar recursion called state evolution  derived heuristically in \cite{Donoho2009} and proved rigorously in \cite{Bayati2011}. State evolution analysis in \cite{Bayati2011} implies that AMP is Bayes-optimal for zero-mean IID sub-Gaussian sensing
matrices when the compression rate is larger than a certain
 threshold \cite{Takeuchi2015}. Spatial coupling \cite{Takeuchi2015, Kudekar2011, Krzakala2012, Donoho2013} is used for the optimality of AMP for any compression rate. %We will not consider spatial coupling in this paper since it is a universal technique for message passing. 

A basic assumption of AMP is that $\bf{A}$ has IID Gaussian entries \cite{Donoho2009,Bayati2011}. For matrices with correlated entries, AMP may perform poorly or even diverge \cite{Manoel2014,Rangan2015,Vila2014}.  It was discovered in \cite{UTAMPa, UTAMPb} that a variant of AMP based on a unitary transformation, called UTAMP, performs well for difficult (e.g. correlated) matrices $\bf{A}$. Independently, orthogonal AMP (OAMP)  was proposed in \cite{Ma2016} for unitarily invariant $\bf{A}$. OAMP is related to a variant of the expectation propagation algorithm \cite{Minka2001}  (called diagonally-restricted expectation consistent inference in \cite{opper2005expectation} or scalar expectation propagation in \cite{Cakmak2018}), as observed in \cite{Takeuchi2017, Rangan2016}. A closely related algorithm, an MMSE-based vector AMP (VAMP) \cite{Rangan2016}, is equivalent to expectation propagation in its diagonally-restricted form \cite{opper2005expectation}.  The accuracy of state evolution for such expectation propagation type algorithms (including VAMP and OAMP) was conjectured in \cite{Ma2016} and proved in \cite{Rangan2016, Takeuchi2017}. The Bayes optimality of OAMP is derived in \cite{Rangan2016, Takeuchi2017, Ma2016} when the compression rate is larger than 
a certain threshold, and the advantages of AMP-type algorithms over conventional turbo receivers \cite{Lei2019TSP, Tuchler2002} are demonstrated in \cite{Lei2019arXiv, MaTWC}.  

The main weakness of OAMP/VAMP is the high-complexity ${\cal O}(M^3 + M^2N)$ incurred by linear MMSE (LMMSE) estimator. Singular-value decomposition (SVD) was used to avoid the high-complexity LMMSE in each iteration \cite{Rangan2016}, but the complexity of the SVD itself is as high as that of the LMMSE estimator. The performance of OAMP/VAMP degrades significantly when the LMMSE estimator is replaced by the low-complexity MF \cite{Ma2016} used in AMP. This limits the application of OAMP/VAMP to large-scale systems for which LMMSE is too complex.
 
In summary, the existing Bayes-optimal AMP-type algorithms are either limited to IID matrices (e.g. AMP) or need high-complexity LMMSE (e.g. OAMP/VAMP). Hence, a low-complexity Bayes-optimal message passing algorithm for unitarily invariant matrices is desired.
 
A long-memory AMP algorithm was originally constructed in
 \cite{Opper2016} to solve the Thouless-Anderson-Palmer
equations for Ising models with general invariant random matrices. The  results in \cite{Opper2016} were rigorously justified via SE in
\cite{Fan2020arxiv}.  Recently, Takeuchi proposed convolutional AMP (CAMP), in which the AMP is modified by replacing the Onsager term with a convolution of all preceding messages \cite{Takeuchi2020CAMP}. The CAMP has low complexity and applies to unitarily invariant matrices. It is proved that the CAMP is Bayes-optimal if it converges to a unique fixed point \cite{Takeuchi2020CAMP}. However, it is found that the CAMP has a low convergence speed and may fail to converge, particularly for matrices with high condition numbers. In addition, a heuristic damping was used to improve the convergence of CAMP. However, the damping is performed on the \emph{a-posteriori} outputs, which breaks orthogonality and the asymptotic Gaussianity of estimation errors \cite{Takeuchi2020CAMP}.

\subsection{Contributions}
To overcome the difficulties in AMP, OAMP/VAMP and CAMP, we propose a memory AMP (MAMP) using a low-complexity long-memory MF. Due to the correlated long memory, stricter orthogonality is required for MAMP to guarantee the asymptotic Gaussianity of estimation errors in MAMP \cite{Takeuchi2017, Takeuchi2020CAMP}. In detail, the step-by-step orthogonalization between current input and output estimation errors in OAMP/VAMP is not sufficient, and instead, the current output estimation error is required to be orthogonal to all preceding input estimation errors. A covariance-matrix state evolution is established for MAMP. Based on state evolution, relaxation parameters and a damping vector, preserving the orthogonality (e.g. the asymptotic Gaussianity of estimation errors), are analytically optimized to guarantee and improve the convergence of MAMP. The main properties of MAMP are summarized as follows.
\begin{itemize}
    \item MAMP has comparable complexity to AMP and much lower complexity than OAMP/VAMP.
    \item MAMP converges to the same fixed point as that of OAMP/VAMP for all unitarily invariant matrices. As a result, it is Bayes-optimal if it has a unique fixed point.   
\end{itemize}  
 
% \begin{definition}[right-unitarily Invariant]\label{Def:unit_inv}
% A random matrix $\bf{A}$ is said to be right-unitarily
% invariant if $\bf{A} \sim  \bf{UAV}$ holds for all deterministic unitary matrices $\bf{U}$ and $\bf{V}$.
% \end{definition} 

\section{Preliminaries} 
%In this section, we first introduce a non-memory iterative process (NMIP) and the orthogonality for NMIP. Then, we simply review  OAMP/VAMP and its properties.

\subsection{Problem Formulation}  
\begin{figure}[t]\vspace{-1mm}
  \centering 
  \includegraphics[width=3.3cm]{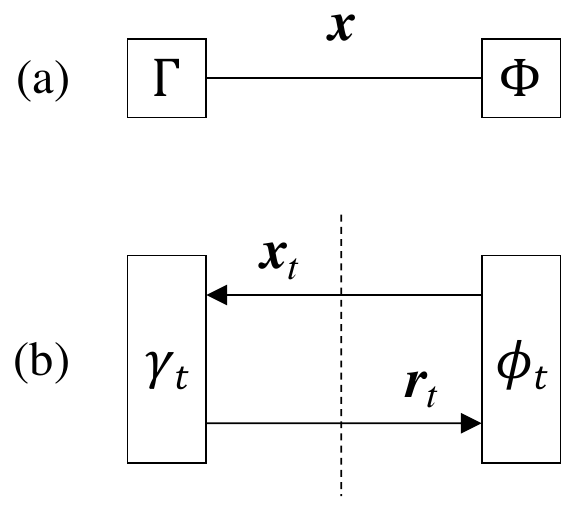}\\ \vspace{-2mm}
  \caption{Graphic illustrations for (a) a system model with two constraints $\Phi$ and $\Gamma$, and (b) a non-memory iterative process (NMIP) involving two local processors $\gamma_t$ and $\phi_t$.}\label{Fig:Model_OAMP} \vspace{-2mm}
\end{figure} 
Fig. \ref{Fig:Model_OAMP}(a) illustrates the system in \eqref{Eqn:linear_system} with two constraints:    
\BE\label{Eqn:unitary_sys}
 \Gamma: \;\;  \bf{y}=\bf{Ax}+\bf{n},\qquad
 \Phi: \;\; x_i\sim P_x, \;\;\forall i. \vspace{-1mm}
\EE
Our aim is to use the AMP-type iterative approach in Fig. \ref{Fig:Model_OAMP}(b) to find an MMSE estimation of $\bf{x}$, i.e., its MSE converges to\vspace{-1mm}
\BE\label{Eqn:post_mean}
{\rm mmse}\{\bf{x}|\bf{y}, \bf{A}, {\Gamma}, {\Phi} \} \equiv \tfrac{1}{N}{\mr{E}}\{\|\hat{\bf{x}}_{\rm post}-{\bf{x}}\|^2\},
\EE
where $\hat{\bf{x}}_{\rm post}\!=\!{\mr{E}}\{\bf{x}|\bf{y}, \bf{A}, {\Gamma}, {\Phi} \}$ is the \textit{a-posteriori} meas of $\bf{x}$.

\begin{definition}[Bayes Optimality]
An iterative approach is said to be Bayes optimal if its MSE converges to the MMSE of the system in \eqref{Eqn:linear_system}.
\end{definition}

\subsection{Assumptions} 
 Let the singular value decomposition of $\bm{A}$ be $\bf{A}\!=\!\bf{U}\bf{\Sigma} \bf{V}$, where $\bf{U}\!\in\! \mathbb{C}^{M\times M}$ and $\bf{V}\!\in\! \mathbb{C}^{N\times N}$ are orthogonal matrices, and $\bf{\Sigma} $ is a diagonal matrix. We assume that $\bf{A}$ is known and is right-unitarily-invariant, i.e., $\bf{U}$, $\bf{V}$ and $\bf{\Sigma}$ are independent, and $\bf{V}$ is Haar distributed. Let $\lambda_t=\tfrac{1}{N}{\rm E}\{{\rm tr}[(\bf{A}\bf{A}^{\rm H})^t]\}$ and ${\lambda}^\dag\equiv[ \lambda_{\max}+ \lambda_{\min}]/2$, where $\lambda_{\min}$ and $\lambda_{\max}$ denote the minimal and maximal eigenvalues of $\bf{A}\bf{A}^{\rm H}$, respectively. We assume that $\lambda_{\min}$, $\lambda_{\max}$ and $\{\lambda_t\}$ are known. This assumption can be relaxed using specific approximations (see \cite{MAMParxiv}).

%  \begin{assumption}\label{Ass:lambda}
%     $\{\lambda_{\min}, \lambda_{\max}, \lambda_t, \forall t\}$ are known. 
%  \end{assumption}
 
 %Assumption \ref{Ass:lambda} is relaxed by specific approximations (see \cite{MAMParxiv}).

\subsection{Non-memory Iterative Process and Orthogonality} 

 {\emph{Non-memory Iterative Process (NMIP):}} Fig. \ref{Fig:Model_OAMP}(b) illustrates an NMIP consisting of a linear estimator (LE) and a non-linear estimator (NLE): Starting with $t=1$,\vspace{-1mm}
\BE\label{Eqn:NMIP}
{\mr{LE:}}\quad \bf{r}_t  = \gamma_t \left(\bf{x}_t\right),\qquad
{\mr{NLE:}} \;\; \bf{x}_{t + 1}  = \phi_t \left( \bf{r}_t \right),\vspace{-1mm}
\EE 
where $\gamma_t(\cdot)$ and $\phi_t(\cdot)$ process the  two  constraints $\Gamma$ and $\Phi$ separately, based only on their current inputs $\bf{x}_t$ and $\bf{r}_t$ respectively. Let\vspace{-1mm}
\BS\label{Eqn:errors}\BE
 \bf{r}_t  = \bf{x} + \bf{g}_t, \qquad
 \bf{x}_t  = \bf{x} + \bf{f}_t, \vspace{-1mm}
\EE
where $\bf{g}_t$ and $\bf{f}_t$ indicate the estimation errors with zero means and variances:
\begin{align}
v_t^\gamma  = \tfrac{1}{N}{\rm E}\{\|\bf{g}_t\|^2 \}, \qquad
v_t^\phi  = \tfrac{1}{N}{\rm E}\{\|\bf{f}_t\|^2 \}. 
\end{align}  \ES
The asymptotic IID Gaussian property of an NMIP was conjectured in \cite{Ma2016} and proved in \cite{Takeuchi2017,Rangan2016}.

\begin{lemma} [Orthogonality and Asymptotic IID Gaussianity]\label{Lem:IIDG}
Assume that  $\bf{A}$ is unitarily invariant with $M, N\!\to\! \infty$ and the following orthogonality holds for all $t\geq1$:
\BE\label{Eqn:error_orth} 
\tfrac{1}{N} \bf{g}_t^{\rm H} \bf{f}_t  \overset{\rm a.s.}{=} 0,\qquad
\tfrac{1}{N} \bf{f}_{t+1}^{\rm H} \bf{g}_t  \overset{\rm a.s.}{=}  0.  \EE
 Then, for Lipschitz-continuous \cite{Berthier2017} $\{{\gamma}_t(\cdot)\}$ and separable-and-Lipschitz-continuous  $\{\phi_t(\cdot)\}$, we have: $\forall t\geq1$, \vspace{-1mm}
\BS\label{Eqn:IIDG} \begin{align}
v_t^{\gamma} &\overset{\rm a.s.}{=} \tfrac{1}{N}{\rm E}\Big\{\|\gamma_t(\bf{x}+\bf{\eta}_t^{\phi})-\bf{x}\|^2\Big\},\\ 
v_{t+1}^{\phi} &\overset{\rm a.s.}{=}  \tfrac{1}{N}{\rm E}\Big\{\|\phi_t(\bf{x}+\bf{\eta}_t^{\gamma})-\bf{x}\|^2\Big\}, 
\end{align}\ES
where $\bf{\eta}_t^{\phi} \sim  \mathcal{CN}(\bf{0},v_t^{\phi} \bf{I})$ and $\bf{\eta}_t^{\gamma} \sim  \mathcal{CN}(\bf{0}, v_t^{\gamma}\bf{I})$ are independent of $\bf{x}$.
\end{lemma} 

%In fact, the sufficient condition of IID Gaussian property for an iterative process is the orthogonality between the current output estimation error and all preceding input estimation errors of each estimator. For NMIP, this condition is relaxed to the orthogonality between current input and output estimation errors, following the generalized Stein's Lemma \cite{Stein1981}. This property has been widely used in the design of orthogonal NMIPs (e.g. OAMP/VAMP and expectation propagation algorithm).

\subsection{Overview of OAMP/VAMP}
%The following OAMP/VAMP  \cite{Ma2016, Rangan2016} is an NMIP that solves the problem in \eqref{Eqn:unitary_sys} with unitarily-invariant matrix.

\emph{OAMP/VAMP \cite{Ma2016, Rangan2016}:} Let $\rho_t  =  \sigma^{2}/ {v_t^\phi}$, $\hat{\phi}_t(\cdot)$ be an MMSE estimator, and $\hat{\gamma}_t(\cdot)$ be an estimator of $\bf{x}$ defined as
\BE\label{Eqn:LMMSE}
\hat{\gamma}_t \left(\bf{x}_t\right) \equiv \bf{A}^{\rm H} \big(\rho_t \bf{I} +   \bf{A}\bf{A}^{\mr H}\big)^{-1}(\bf{y}-\bf{A} \bf{x}_t ).
\EE 
An OAMP/VAMP is then defined as: Starting with $t=1$, $v_1^\phi=1$ and $\bf{x}_1=\bf{0}$,\vspace{-1mm}
\BS\label{Eqn:OAMP/VAMP}\begin{align}
\!\!\!\!{\mr{LE:}} \quad  \bf{r}_t  =  \gamma_t \left(\bf{x}_t\right)&\equiv  \tfrac{1}{ {\epsilon}^\gamma_t}     \hat{\gamma}_t \left(\bf{x}_t\right) + \bf{x}_t , \label{Eqn:OAMP/VAMP_LE}\\
\!\!\!\!{\mr{NLE:}} \; \bf{x}_{t+1}  \!=\! \phi_t \left( \bf{r}_t \right)&\!\equiv\!  \tfrac{1}{\epsilon^\phi_{t+1}} \!\big[  \hat{\phi}_t(\bf{r}_t) \!+\! (\epsilon^\phi_{t+1}\!-\!1) \bf{r}_t \big],\label{Eqn:OAMP/VAMP_NLE}
\end{align}
where \vspace{-3mm}
\begin{align}
  {\epsilon}_t^\gamma &= \tfrac{1}{N}{\mr {tr}} \big\{\bf{A}^{\rm H}\big(\rho_t\bf{I} +   \bf{A}\bf{A}^{\mr H}\big)^{\!-1}\!\!\bf{A}\big\}, \\ v^\gamma_t &=\gamma_{\mr{SE}}(v_t^\phi) \equiv {v}^\phi_t  [({\epsilon}_t^\gamma) ^{-1} -1],  \label{Eqn:OAMP_vara} \end{align}\begin{align}
 \epsilon^\phi_{t+1} &= 1-\tfrac{1}{N v^\gamma_t}\|\hat{\phi}_t(\bf{x}+\sqrt{v^\gamma_t} \bf{\eta} ) -\bf{x}\|^2, \\ {v}^\phi_{t+1} &=  \phi_{\mr{SE}}(v^\gamma_t) \equiv v^\gamma_t [(\epsilon^\phi_{t+1}) ^{-1} -1],\label{Eqn:OAMP_varb}
\end{align}\ES
and $\bf{\eta}\sim {\cal{CN}}(\bf{0}, \bf{I})$ is independent of $\bf{x}$.

We assume that $\hat{\phi}_t(\cdot)$ is an MMSE estimator given by
\BE
 \hat{\phi}_t(\bf{r}^{t})\equiv \mr{E}\{\bf{x}|\bf{r}^{t}\}.
\EE 
It was proved in  \cite{Takeuchi2017} that OAMP/VAMP satisfies the orthogonality in \eqref{Eqn:error_orth}. Hence, the IID Gaussian property in \eqref{Eqn:IIDG} holds for OAMP/VAMP.

% \emph{State Evolution:} The iterative performance can be tracked by the following state evolution: Starting with $t=1$ and $v^\phi_1=1$,\vspace{-2mm}
% \BS\begin{align}
% {\mr{LE:}} \qquad  v^\gamma_t &=\gamma_{\mr{SE}}(v^\phi_t),    \\
% {\mr{NLE:}} \quad   {v}^\phi_{t+1}  &=  \phi_{\mr{SE}}(v^\gamma_t),
% \end{align}\ES 
% where $\gamma_{\mr{SE}}(\cdot)$ and $\phi_{\mr{SE}}(\cdot)$ are defined in \eqref{Eqn:OAMP_vara} and \eqref{Eqn:OAMP_varb}, respectively.

\begin{lemma}[Bayes Optimality \cite{Barbier2018b, Tulino2013,Ma2016,Takeda2006}]\label{Lem:optimality}
Assume that $M,\!N\!\to\!\infty$ with a fixed $\delta\!=\!\!M\!/\!N$, and OAMP/VAMP satisfies the unique fixed point condition. Then, OAMP/VAMP is Bayes optimal for right-unitarily-invariant matrices.
\end{lemma}

\emph{Complexity:} The NLE in OAMP/VAMP is a symbol-by-symbol estimator, whose time complexity is as low as ${\cal O}(N)$. The complexity of OAMP/VAMP is dominated by LMMSE-LE, which costs $\mathcal{O}(M^2N\!+\!M^3)$ time complexity per iteration for matrix multiplication and matrix inversion. %Therefore, to reduce the complexity, it is desired to design a low-complexity Bayes-optimal LE for the message passing algorithm. 

\section{Memory AMP}\label{Sec:MAMP}
%In this section, we first introduce the memory iterative process (MIP) and the orthogonality for MIP. Then, construct a memory AMP (MAMP) using a long-memory MF-LE. Interestingly, with the assistance of memory, MAMP converges to the Bayes-optimal OAMP/VAMP even a low-complexity MF is used. 

 \subsection{Memory Iterative Process and Orthogonality}
\begin{figure}[t]\vspace{-1mm}
  \centering 
  \includegraphics[width=3.5cm]{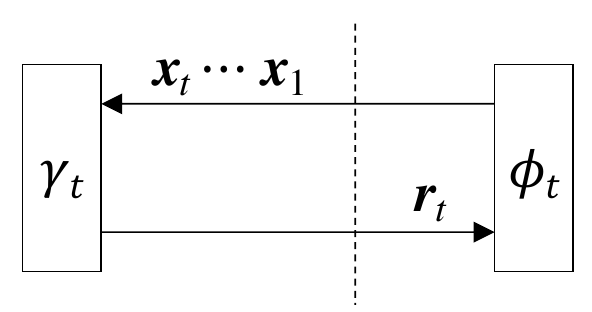}\\\vspace{-2mm} 
  \caption{Graphic illustration for a long-memory iterative process (MIP).}\label{Fig:MAMP}
\end{figure}
{\emph{Memory  Iterative Process (MIP):}} Fig. \ref{Fig:MAMP} illustrates an MIP based on a long-memory linear estimator (LMLE) and a non-linear estimator (NLE) defined as:  Starting with $t=1$, 
\BS\label{Eqn:MIP}\begin{alignat}{2}
{\rm LMIE:}&& \quad \quad  \bf{r}_t &= \gamma_t \left(\bf{x}_1,\cdots \bf{x}_t\right),\\
{\rm NLE:}&& \quad   \bf{x}_{t + 1} &= \phi_t \left( \bf{r}_t \right).
\end{alignat}\ES

We call \eqref{Eqn:MIP} MIP since $\gamma_t(\cdot)$ contains long memory $ \{\bf{x}_i, i<t\}$. It should be emphasized that the step-by-step orthogonalization between current input and output estimation errors is not sufficient to guarantee the asymptotic IID Gaussianity for MIP. Thus, a stricter orthogonality is required, i.e., the estimation error of $\gamma_t(\cdot)$ is required to be orthogonal to all preceding estimation errors \cite{Takeuchi2020CAMP, Takeuchi2019AMP}.

\begin{lemma} [Orthogonality and Asymptotic IID Gaussianity \cite{Takeuchi2020CAMP, Takeuchi2019AMP}]\label{Lem:IIDG_MIP}
Assume that $\bf{A}$ is unitarily invariant with $M, N\!\to\! \infty$ and the following orthogonality holds for all $1\le t'\leq t$:
\BE\label{Eqn:error_orth_MIP} 
    \tfrac{1}{N} \bf{g}_t^{\rm H}\bf{f}_{t'}   \overset{\rm a.s.}{=}  0,\qquad
 \tfrac{1}{N}\bf{f}_{t+1}^{\rm H}\bf{g}_t     \overset{\rm a.s.}{=} 0. 
 \EE
Then, for Lipschitz-continuous \cite{Berthier2017} $\{{\gamma}_t(\cdot)\}$ and separable-and-Lipschitz-continuous  $\{\phi_t(\cdot)\}$,  we have: $\forall 1\!\le \!t'\!\leq\! t$, \vspace{-1mm}
\BS\label{Eqn:IIDG_MIP}  \begin{align}
&v_{t,t'}^{\gamma} \!\!\overset{\rm a.s.}{=} \!\!\tfrac{1}{N}{\rm E}\big\{\big[\gamma_t\big(\bf{x}+\bf{\eta}_1^{\phi}, \dots, \bf{x}+\bf{\eta}_t^{\phi}\big)-\bf{x}\big]^{\rm H}\nonumber\\
& \hspace{2cm}\cdot\big[\gamma_{t'}\big(\bf{x}+\bf{\eta}_1^{\phi}, \dots, \bf{x}+\bf{\eta}_{t'}^{\phi}\big)-\bf{x}\big]\big\},\\ 
&v_{t+1,{t'}\!+1}^{\phi} \!\!\overset{\rm a.s.}{=} \!\! 
\tfrac{1}{N}{\rm E}\big\{\! \big[\phi_t\big(\bf{x}\!+\!\bf{\eta}_t^{\gamma}\big)\!-\!\bf{x}\big]^{\!\rm H}\big[\phi_{t'}\big(\bf{x}\!+\!\bf{\eta}_{t'}^{\gamma}\big)\!-\!\bf{x}\big]\! \big\}\!,  
\end{align}\ES 
where $\bf{\eta}_t^{\phi}\!\!\sim\!\mathcal{CN}(\bf{0},v_{t,t}^{\phi}\bf{I})$ with ${\rm E}\{\bf{\eta}_\tau^{\phi}(\bf{\eta}_{\tau'}^{\phi})^{\rm H}\}\!=\!v_{\tau,\tau'}^{\phi}\bf{I}$ and $\bf{\eta}_t^{\gamma}\!\!\sim\!\mathcal{CN}(\bf{0},v_{t,t}^{\gamma}\bf{I})$ with ${\rm E}\{\bf{\eta}_\tau^{\gamma}(\bf{\eta}_{\tau'}^{\gamma})^{\rm H}\}\!=\!v_{\tau,\tau'}^{\gamma}\bf{I}$. Besides, $\{\bf{\eta}_t^{\phi}\}$ and $\bf{\eta}_t^{\gamma}\}$ are independent of $\bf{x}$.  
\end{lemma} 

% \begin{lemma} [Orthogonality and Asymptotic IID Gaussianity \cite{Takeuchi2020CAMP, Takeuchi2019AMP } ]\label{Lem:IIDG_MIP}
% Assume that $M, N\!\to\! \infty$ with fixed $M/N$ and the following orthogonality holds for $1\le t'\leq t$:
%  \begin{align}\label{Eqn:error_orth_MIP}
% {\rm E}\{\bf{g}_t^{\rm H}\bf{f}_{t'} \}    = 0,\qquad
% {\rm E}\{\bf{f}_{t+1}^{\rm H}\bf{g}_t \}   = 0.
% \end{align}   
% Then, for pseudo-Lipschitz functions \cite{Berthier2017}  $\{\gamma_t(\cdot)\}$ and  $\{\phi_t(\cdot)\}$, we have\vspace{-1mm}
% \BE\label{Eqn:IIDG_MIP} 
% \bf{r}_t   = \bf{x}+ \bf{\eta}_t,\vspace{-1mm}
% \EE
% where $\bf{\eta}_t$ is IID Gaussian and is independent of $\bf{x}$. More specifically, any finite rows of $[\bf{\eta}_1\cdots\bf{\eta}_t]$ are IID with Gaussian distribution.   
% \end{lemma}   

\subsection{Memory AMP (MAMP)}   
 \emph{Memory AMP:} Let $\bf{B}  = \lambda^\dag\bf{I} - \bf{A}\bf{A}^{\mr H}$, and consider  \vspace{-1mm}
 \BE\label{Eqn:MLEa}
 \hat{\bf{r}}_{t}= 
 \theta_t  \bf{B} \hat{\bf{r}}_{t-1} +   \xi_t(\bf{y} - \bf{A}\bf{x}_t).  \vspace{-1mm}
 \EE
 An MAMP process is defined as: Starting with $t\!=\!\!1$ and ${\bf{x}}_{1}\!\!=\!\hat{\bf{r}}_{0}\!=\!\bf{0}$, \vspace{-1mm}
\BS\label{Eqn:MAMP}
\begin{alignat}{2}
  \bf{r}_t &=\gamma_t \left(\bf{x}_1,\cdots \bf{x}_t\right)\equiv  \tfrac{1}{{\varepsilon}_t}\big( \bf{A}^{\mr H}\hat{\bf{r}}_{t} + \textstyle\sum_{i=1}^t  p_{t i}\bf{x}_i \big), \label{Eqn:MLE}\\%
  \bf{x}_{t + 1} &= \bar{\phi}_t(\bf{r}_t) =\scaleto{\zeta}{8pt}_{tl_t} \phi_t \left( \bf{r}_t \right) + \textstyle\sum_{i=1}^{l_t-1}{\scaleto{\zeta}{8pt}}_{ti}\bf{x}_{t-l_t+1+i},\label{Eqn:NLE}
\end{alignat}\ES	 
where $\phi_t(\cdot)$ is the same as that in OAMP/VAMP (see \eqref{Eqn:OAMP/VAMP}).

The following are intuitions of MAMP.  
\begin{itemize} 
    \item In LMIE, all preceding messages are utilized in $\textstyle\sum_{i=1}^t  p_{t i}\bf{x}_i$ to guarantee the orthogonality in \eqref{Eqn:error_orth_MIP}. In NLE, at most $l_t-1$ preceding messages are utilized in $\textstyle\sum_{i=1}^{l_t-1}\scaleto{\zeta}{8pt}_{ti}\bf{x}_{t-l_t+1+i}$ (e.g. damping) to guarantee and improve the convergence of MAMP.
    \item $\bf{B}  = \lambda^\dag\bf{I} - \bf{A}\bf{A}^{\mr H}$ ensures that MAMP has the same fixed point as OAMP/VAMP.
    \item $\{{\varepsilon}_t\}$ and $\{ p_{t i}\}$, given in Subsection \ref{Sec:orth} (see \eqref{Eqn:orth_parameters}), guarantee the orthogonality in \eqref{Eqn:error_orth_MIP} (see Theorem \ref{The:IIDG_MIP}).  
    \item $\{\theta_t\}$ and $\{\xi_t\}$ improve the convergence speed of MAMP. The optimizations of $ \theta_t$ and $\xi_t$ are given in Subsections \ref{Sec:theta_opt} (see \eqref{Eqn:opt_xi}) and \ref{Sec:xi} (see \eqref{Eqn:theta_opt}), respectively.   
    \item $\scaleto{\bf{\zeta}}{8pt}_t=[{\scaleto{\zeta}{8pt}}_{t1}, \cdots, {\scaleto{\zeta}{8pt}}_{tl_t}]^{\rm T}$, optimized in Subsection \ref{Sec:vdelta} (see \eqref{Eqn:vdelta_opt}), is a damping vector with $\textstyle\sum_{i=1}^{l_t} {\scaleto{\zeta}{8pt}}_{ti} =1$. In particular, no damping is applied for $\scaleto{\bf{\zeta}}{8pt}_t=[0,\cdots,0, 1]^{\rm T}$. We set $l_t=\min\{L, t\!+\!1\}$, where $L$ is the maximum damping length (in general, $L =3$). Damping guarantees and improves the convergence of MAMP.   
\end{itemize}

We call \eqref{Eqn:MAMP} memory AMP as it involves a long memory $ \{\bf{x}_i, i<t\}$ at LMIE that is different from the non-memory LE in OAMP/VAMP. Matrix-vector multiplications instead of matrix inverse are involved. Thus, the complexity of MAMP is comparable to AMP, i.e., as low as ${\cal O}(MN)$ per iteration.

\section{Main Properties of MAMP} 
%In this section, we provide some important properties of MAMP such as optimality, orthogonality and asymptotic IID Gaussianity, based on which we derive the state evolution of MAMP.
 
\subsection{Orthogonality and Asymptotic IID Gaussianity} \label{Sec:orth}
Let $\bf{W}_t = \bf{A}^{\rm H}\bf{B}^{t}\bf{A}$. For $t\ge 0$, we define   
\BS\label{Eqn:sym_defs}\begin{align}
 b_t &\equiv \tfrac{1}{N}{\rm tr}\{\bf{B}^t\} = \textstyle\sum_{i=0}^{t} \binom{t}{i} (-1)^i(\lambda^\dag)^{t-i}\lambda_i,\\  %\!\! 
   w_t &\equiv  \tfrac{1}{N}{\rm tr}\{\bf{W}_t\}= \lambda^\dag b_{t}- b_{t+1}.\label{Eqn:b_w}
\end{align}  \ES 
For $1\le i\le t$,   
\begin{align}\label{Eqn:orth_parameters}
 \!\! \vartheta_{t i}   \equiv \xi_i  \!\textstyle\prod_{\tau=i+1}^t \! \theta_\tau, \;\;\;  p_{t i}\equiv  \vartheta_{t i}   w_{t-i}, \;\;\;  {\varepsilon}_t  = \textstyle\sum_{i=1}^t  p_{t i}.
\end{align}  
Furthermore, $\vartheta_{t i}=1$ if $i>t$.      
 
\begin{proposition}\label{Pro:MLE-e}  
The $\{{\bf{r}}_t \}$ in \eqref{Eqn:MAMP} and the corresponding errors can be expanded to
\BS\begin{align}
 \!\! {\bf{r}}_t  \! \!=\! \!\tfrac{1}{ {\varepsilon}_t }\! \big[\bf{F}_t\bf{y} \!+\!\! \textstyle\sum_{i=1}^t \!\!\bf{H}_{t i}\bf{x}_i  \big], \;\,
\bf{g}_t   \!\!= \! \! \tfrac{1}{ {\varepsilon}_t }\! \big(\bf{F}_t{\bf{n}} \!+\!\! \textstyle\sum_{i=1}^t \!\!\bf{H}_{t i}\bf{f}_i \big)\!, 
\end{align}
where 
\begin{align}
\bf{F}_t \! \equiv\!   \textstyle\sum_{i=1}^{t} \vartheta_{t i}\bf{A}^{\mr H}  \bf{B}^{t-i},\quad
\bf{H}_{t i} \! \equiv \! \vartheta_{t i}(  w_{t-i}\bf{I}\! -\! \bf{W}_{t-i}). 
\end{align}\ES
\end{proposition} 

The following is based on Lemma \ref{Lem:IIDG_MIP} and Proposition \ref{Pro:MLE-e}.
 
\begin{theorem}[Orthogonality and Asymptotic IID Gaussianity]\label{The:IIDG_MIP}
Assume that $\bf{A}$ is right-unitarily-invariant with $M, N\!\to\! \infty$. The  orthogonality in \eqref{Eqn:error_orth_MIP} holds for MAMP. Therefore, the IID Gaussianity in \eqref{Eqn:IIDG_MIP} holds for MAMP.
\end{theorem}

% \begin{IEEEproof}
% See APPENDIX \ref{APP:IIDG_MIP}.  
% \end{IEEEproof} 

% Using Theorem \ref{The:IIDG_MIP}, we can track the performance of MAMP using the state evolution discussed in the following subsection.

\subsection{State Evolution}\label{Sec:SE}
Using the IID Gaussian property in Theorem \ref{The:IIDG_MIP}, we establish a state evolution for the dynamics of the MSE of MAMP. The main challenge is the correlation between the long-memory inputs of LMIE. It requires a  covariance-matrix state evolution to track the dynamics of MSE. 

Define the covariance matrices as follows:\vspace{-1mm}
\BS\BE
   \bf{V}_t^{\gamma}  =[v^{\gamma}_{ij}]_{t\times t},\qquad
     \bf{V}_t^{\bar{\phi}} =[v^{\bar{\phi}}_{ij}]_{t\times t},\vspace{-1mm}
\EE  
where\vspace{-1mm}
\begin{align}
   v^{\gamma}_{tt'} \equiv \tfrac{1}{N}{\mr E} \{\bf{g}_t^{\mr H} \bf{g}_{t'} \},\qquad
   v^{\bar{\phi}}_{tt'}  \equiv \tfrac{1}{N}{\mr E} \{\bf{f}_t^{\mr H} \bf{f}_{t'}\}.
\end{align} 
\ES   

\begin{proposition}[State Evolution]\label{Pro:SE_var}
The covariance matrices of MAMP can be tracked by the following state evolution: Starting with $v_{11}^{\phi}=1$,
\BE\label{SE_MAMP} 
{\rm LMIE:}  \; \bf{V}_t^{\gamma} = \gamma_{\mr{SE}}(\bf{V}_t^{\bar{\phi}}), \quad
{\rm NLE:} \;  \bf{V}_{t+1}^{\bar{\phi}}   =\bar{\phi}_{\mr{SE}}(\bf{V}_t^{\gamma}).  
 \EE
The details of $\gamma_{\mr{SE}}(\cdot)$ and $\bar{\phi}_{\mr{SE}}(\cdot)$ are provided in \cite{MAMParxiv}. 
\end{proposition}

\subsection{Convergence and Bayes Optimality}  
The following theorem gives the convergence and Bayes optimality of the optimized MAMP. The proof is provided in Appendix F in \cite{MAMParxiv}.
 
\begin{theorem}[Convergence and Bayes Optimality]\label{The:Conv_MOAMP/VAMP}
Assume that $M,N\to\infty$ with a fixed $\delta=M/N$ and $\bf{A}$ is right-unitarily-invariant. The MAMP with optimized $\{\theta_t, \bf{\zeta}_t\}$ (see Section \ref{Sec:optimization}) converges to the same fixed point as OAMP/VAMP, i.e., it is Bayes optimal if it has a unique fixed point.
\end{theorem}

\subsection{Complexity Comparison}  
Table \ref{Tab:complexity} compares the time and space complexity of MAMP, CAMP, AMP and OAMP/VAMP, where $T$ is the number of iterations. MAMP and CAMP have the similar time and space complexity.  OAMP/VAMP has higher complexity than AMP, CAMP and MAMP, while MAMP and CAMP have comparable complexity to AMP for $T \ll N$. For more details, refer to Section IV-D in \cite{MAMParxiv}.
 \begin{table}[hbt]\vspace{-3mm}
\renewcommand{\arraystretch}{1.4} 
\caption{ \scriptsize Time and Space Complexity Comparison Between AMP-Type Algorithms}
\label{Tab:complexity}
\centering \scriptsize \setlength{\tabcolsep}{1.0mm}{
\begin{tabular}{|c|c|c|}
\hline
Algorithms & Time complexity & Space complexity   \\
\hline
AMP\cite{Donoho2009} &  ${\cal O}(MNT )$ & ${\cal O}(MN) $   \\ 
 \hline 
 \tabincell{c}{OAMP/VAMP\cite{Rangan2016, Ma2016}\vspace{-0.1cm}\\ (SVD)}  &  ${\cal O}(M^2N\!+\!MNT)$ & ${\cal O}(N^2\!+\!MN)$   \\
 \hline
 \tabincell{c}{OAMP/VAMP\cite{Rangan2016, Ma2016}\vspace{-0.1cm}\\ (matrix inverse)}  &  ${\cal O}\big((M^2N\!+\!M^3)T\big)$ & ${\cal O}(MN)$   \\
\hline
CAMP \cite{Takeuchi2020CAMP}  & ${\cal O}(MNT\!+\!MT^2\!+\!T^4)$ & ${\cal O}(MN\!+\!MT \!+\!T^2) $      \\
\hline 
 MAMP   & ${\cal O}\big(M\!NT\!\!+\!(\!N\!\!+\!\!M\!)T^2\!\!+\!\!T^3\big)$ & ${\cal O}\big(M\!N\!\!+\!(\!N\!\!+\!\!M\!)T \!\!+\!\!T^2\big) $    \\
\hline
\end{tabular}}
\end{table}

\section{Parameter Optimization}\label{Sec:optimization}
The parameters $\{\theta_t, \xi_t, \scaleto{\bf{\zeta}}{8pt}_t\}$ are optimized step-by-step for each iteration assuming that the parameters $\{\theta_{t'}, \xi_{t'}, \scaleto{\bf{\zeta}}{8pt}_{t'}, {t'}\leq t-1\}$ in previous iterations are fixed. More specifically, we first optimize $\theta_t$. Then, given  $\theta_t$, we optimize $\xi_t$. Finally, given $\theta_t$ and $\xi$, we optimize $\scaleto{\bf{\zeta}}{8pt}_t$.

\subsection{Optimization of \texorpdfstring{$\theta_t$}{TEXT}} \label{Sec:theta_opt}
The optimal $\theta_t$  is given by  
 \BE\label{Eqn:theta_opt}
 \theta_t =(\lambda^\dag+\sigma^2/v_{tt}^\phi)^{-1},
 \EE
which minimizes the spectral radius of $\theta_t\bf{B}$. From \eqref{Eqn:theta_opt}, the spectral radius of $\theta_t\bf{B}$ satisfies
\BE
\rho(\theta_t\bf{B}) =   \tfrac{\lambda_{\max}- \lambda_{\min}}{ \lambda_{\max}+ \lambda_{\min} +2\rho_t} <1. 
\EE 
That is, the convergence condition is satisfied. In addition,  \eqref{Eqn:theta_opt} also optimizes the convergence speed of MAMP. 

\subsection{Optimization of \texorpdfstring{$\xi_t$}{TEXT}} \label{Sec:xi}
 
\begin{proposition}\label{Pro:xi_opt}
Fixed $\theta_t$, the optimal $\xi_t$ that minimizes $v_{tt}^{\gamma}$ is given by $\xi_1^{\rm opt}=1$ and  for $ t\ge 2$,
 \BE\label{Eqn:opt_xi}
\xi_t^{\rm opt}= \frac{c_{t2}c_{t0}+c_{t3}}{c_{t1}c_{t0}+c_{t2}},
\EE  
where 
\BS\label{Eqn:cs}\begin{align}
c_{t0} &= \textstyle\sum_{i=1}^{t-1} p_{t i}/ w_0,\qquad\qquad
c_{t1} =\sigma^2  w_0+ v_{{t}{t}}^{\phi}\bar{ w}_{00},\\
c_{t2}&= - \textstyle{\sum}_{i=1}^{t-1} \vartheta_{t i}   (\sigma^2 w_{t-i}+ v_{{t}{i}}^{\phi}\bar{ w}_{0t-i} ), \\
c_{t3}&= \textstyle{\sum}_{i=1}^{t-1}\textstyle{\sum}_{j=1}^{t-1}  \vartheta_{t i}\vartheta_{tj}\big(\sigma^2  w_{2t-i-j} + v^{\phi}_{ij}\bar{ w}_{t-it-j}\big).  
\end{align}\ES
\end{proposition}

%When $c_{t1}c_{t0}+c_{t2}$ is equal or close  to zero, to avoid overflow in simulation, we can set $\xi_t^{\rm opt}={\rm sign}\{\xi_t^{\rm opt}\}\cdot C_{\rm max}$ if $|\xi_t^{\rm opt}|>C_{\rm max}$, where $C_{\rm max}$ is a sufficiently large number. 

 \subsection{Optimization of \texorpdfstring{$\scaleto{\bf{\zeta}}{8pt}_t$}{TEXT}} \label{Sec:vdelta}      
Let $\bf{\mathcal{V}}_{t+1}^\phi$ be the covariance matrix of the input errors of $\bar{\phi}_t$, i.e. $\{\bf{f}_{t-l_t+2},\cdots, \bf{f}_{t}, \phi_{t}(\bf{r}_{t})-\bf{x} \}$. 

\begin{proposition}[Optimal damping]\label{Pro:vdamp}
Fixed $\theta_t$ and $\xi_t$, the optimal $\scaleto{\bf{\zeta}}{8pt}_t$ that minimizes $v_{t+1t+1}^{\bar{\phi}}$ is given by
\BE\label{Eqn:vdelta_opt}
   \scaleto{\bf{\zeta}}{8pt}_t^{\rm opt}  = \tfrac{  [\bf{\mathcal{V}}_{t+1}^\phi]^{-1} \bf{1}}{\bf{1}^{\rm \!T} [\bf{\mathcal{V}}_{t+1}^\phi]^{-1}\bf{1}}.
\EE
\end{proposition}

It is easy to see that the MSE of current iteration with optimized damping is not worse than that of the previous iteration, which is a special case of $\scaleto{\bf{\zeta}}{8pt}_t=[0, \cdots,1, 0]^{\rm T}$. That is, the MSE of MAMP with optimized damping is monotonically decreasing in the iterations. %Therefore, the optimized $\scaleto{\bf{\zeta}}{8pt}_t$ guarantees the convergence of MAMP as well as improves the convergence speed of MAMP.
 
%The calculation of $\{\scaleto{\bf{\zeta}}{8pt}_t^{\rm opt}\}$ in \eqref{Eqn:vdelta_opt} costs time complexity of ${\cal O}(L^3T)$ due to the matrix inverse $[\bf{\mathcal{V}}_{t+1}^\phi]^{-1}$, where $L$ is the length of damping vector and $T$ the number of iterations. Thus, the time complexity of MAMP is changed to ${\cal O}\big(MNT\!+\!(N\!+\!M)T^2\!+\!T^3\!+\!L^3T\big)$, where ${\cal O}(L^3T)$ is negligible when $L\ll T\ll M$. In general, the best choice of $L$ is less than or equal to 3 (see the simulation results in Section \ref{Sec:Simulation}).
 
%Proposition \ref{Pro:vdamp} is based on the condition $v^{\phi}_{t+1t+1}  - 2 v^{\phi}_{t+1t}  +v^{\bar{\phi}}_{tt}=\tfrac{1}{N}{\rm E}\{\|\phi_{t}(\bf{r}_{t})-\bf{x}_{t}\|^2\}\ge 0$, which may not hold due to the inaccuracy of $\{v^{\phi}_{tt'}\}$ in the simulation. To avoid this problem, we can set $\zeta=1$, when $v^{\phi}_{t+1t+1}  - 2 v^{\phi}_{t+1t}  + v^{\bar{\phi}}_{tt}<0$ happens in the simulation. 
  
  \section{Simulation Results} \label{Sec:Simulation}
We study a compressed sensing problem where $\bf{x}$ follows a symbol-wise Bernoulli-Gaussian distribution, i.e. $\forall i$,
 \BE
 x_i\sim \left\{\begin{array}{ll}
      0,&  {\rm probability} = 1- \mu  \\
      {\cal N} (0, \mu^{-1}),&  {\rm probability} = \mu
 \end{array}\right..
 \EE
The variance of $x_i$ is normalized to 1. The signal-to-noise ratio (SNR) is defined as ${\rm SNR}=1/\sigma^2$.
 
 Let the SVD of $\bm{A}$ be $\bm{A}=\bm{U \Sigma V}$. The system model in \eqref{Eqn:linear_system} can be rewritten as\cite{Ma2016,Rangan2016}:
\BE
{\bf{y}} = {\bm{U \Sigma V}}\bf{x} + {\bf{n}}.
\EE
Note that $\bm{U}^{\mr{H}}\bm{n}$ has the same distribution as $\bm{n}$. Thus, we can assume $\bm{U} =\bm{I}$ without loss of generality. To reduce the calculation complexity of OAMP/VAMP, we approximate a large random unitary matrix by ${\bf{V}} = {\bf{\Pi F}}$, where $\bm{\Pi}$ is a random permutation matrix and $\bm{F}$ is a discrete Fourier transform (DFT) matrix. Note that all the algorithms involved here admit fast implementation for this matrix model. The eigenvalues $\{d_i\}$ are generated as: $d_i/d_{i+1}=\kappa^{1/
J}$ for $i = 1,\ldots, J-1$ and $\sum_{i=1}^Jd_i^2=N$, where $J=\min\{M, N\}$. Here, $\kappa\ge1$ controls the condition number of $\bm{A}$. Note that MAMP does not require the SVD structure of $\bf{A}$. MAMP only needs the right-unitarily invariance of $\bf{A}$.\vspace{-2mm}

\subsection{Influence of Relaxation Parameters and Damping}
\begin{figure}[t]
  \centering 
  \includegraphics[width=.33\textwidth]{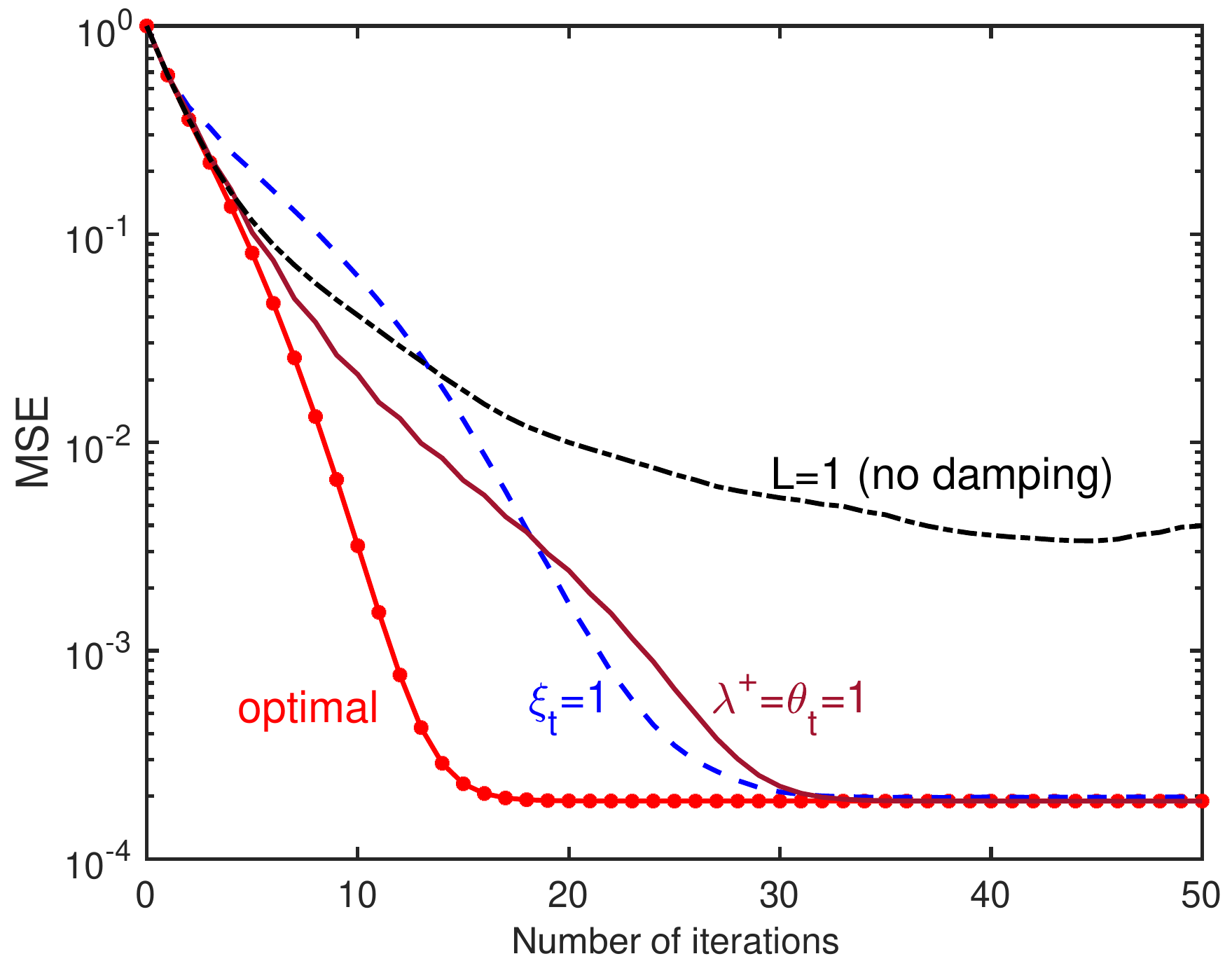}\\
  \caption{MSE versus the number of iterations for MAMP with different parameters $\{\lambda^\dag, \theta_t, \xi_t, L\}$. $M=4096, N=8192$, $\mu=0.1$,  $\kappa=10$  and ${\rm SNR}=30$ dB. ``optimal" denotes MAMP with optimized $\{\lambda^\dag, \theta_t, \xi_t\}$ and $L=3$ (damping length). The other curves denote MAMP with the same parameters as the ``optimal" except the one marked on each curve.}\label{Fig:MAMP_parameters} \vspace{-0.2cm}
\end{figure}

Fig.~\ref{Fig:MAMP_parameters} shows the influence of the relaxation parameters $\{\lambda^\dag, \theta_t, \xi_t\}$ and damping. Without damping (e.g. $L=1$) the convergence of MAMP is not guaranteed. In addition, the optimization of $\{\lambda^\dag, \theta_t, \xi_t\}$ has significant improvement in the MSE of MAMP. That is
\begin{itemize}
    \item[(i)] damping guarantees the convergence of MAMP, and
    \item[(ii)] the relaxation parameters $\{\lambda^\dag, \theta_t, \xi_t\}$ do not change the  fixed point of MAMP, but they can be optimized to improve the convergence speed.
\end{itemize}
  
\subsection{Comparison with AMP and CAMP}   
\begin{figure}[t]
\centering
\includegraphics[width=.33\textwidth, height=.2\textheight]{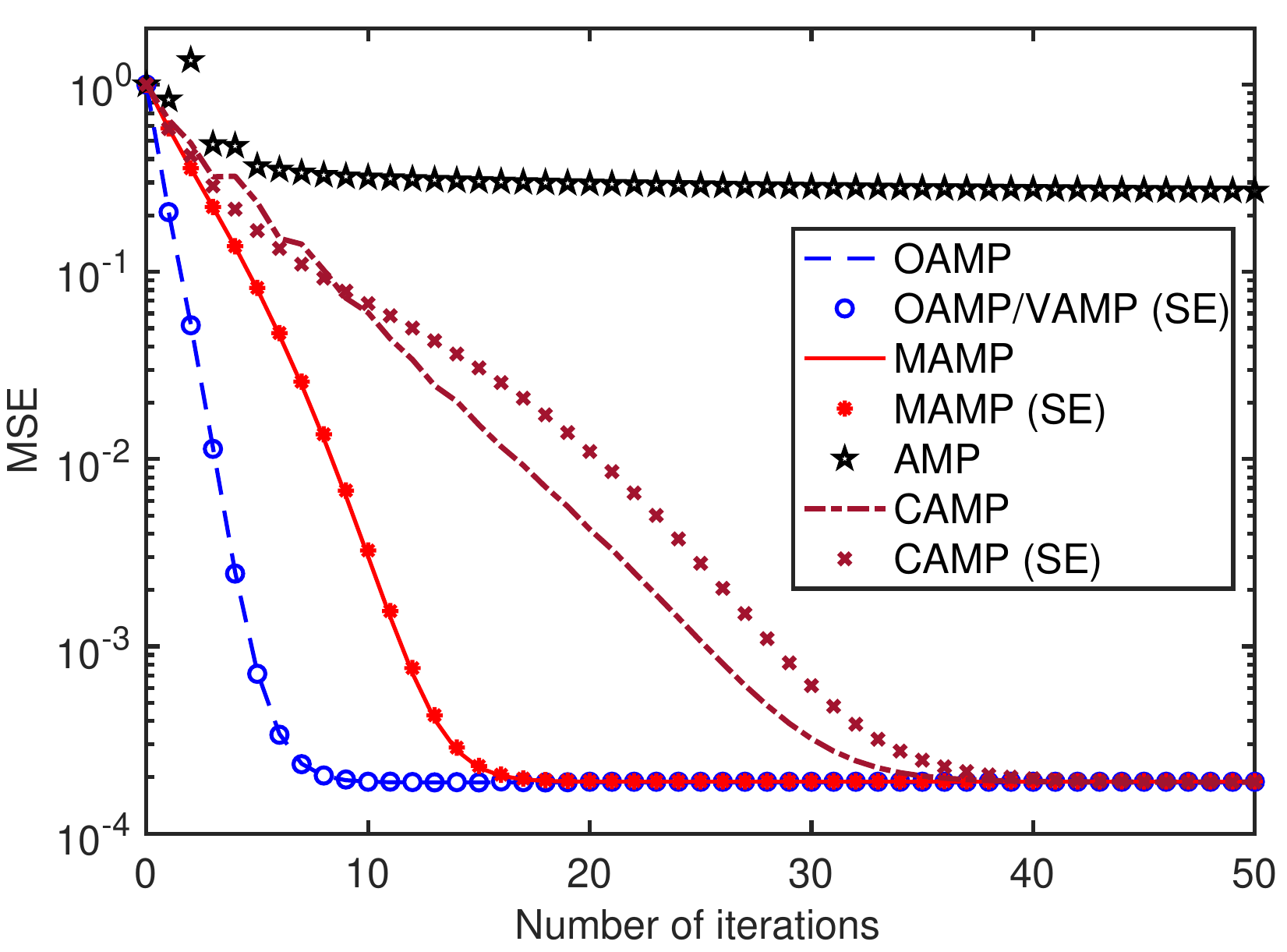}   
\caption{MSE versus the number of iterations for AMP, CAMP, OMAP/VAMP and MAMP. $M=8192, N=16384$, $\mu=0.1$,  $\kappa=10$, $L= 3$  and ${\rm SNR}=30$ dB. The curves of AMP and CAMP  are from Fig. 2 in \cite{Takeuchi2020CAMP}.}
\label{Fig:CAMP_MAMP} %\vspace{-2mm}
\end{figure}

Fig.~\ref{Fig:CAMP_MAMP} shows MSE versus the number of iterations for AMP, CAMP, OMAP/VAMP and MAMP. To improve the convergence, both AMP and CAMP are damped. As can be seen, for an ill-conditioned matrix with $\kappa=10$, the MSE performance of AMP is poor. CAMP converges to the same performance as that of OAMP/VAMP. However, the state evolution (SE) of CAMP is inaccurate since damping is made on the \emph{a-posteriori} outputs, which breaks the Gaussianity of the estimation errors. MAMP converges faster than CAMP to OAMP/VAMP. Furthermore, the state evolution of MAMP is accurate since damping is made on the \emph{orthogonal} outputs, which preserves the Gaussianity of the estimation errors.

\subsection{Influence of High Condition Number and Damping Length} 
\begin{figure}[t]
\centering 
\includegraphics[width=.33\textwidth]{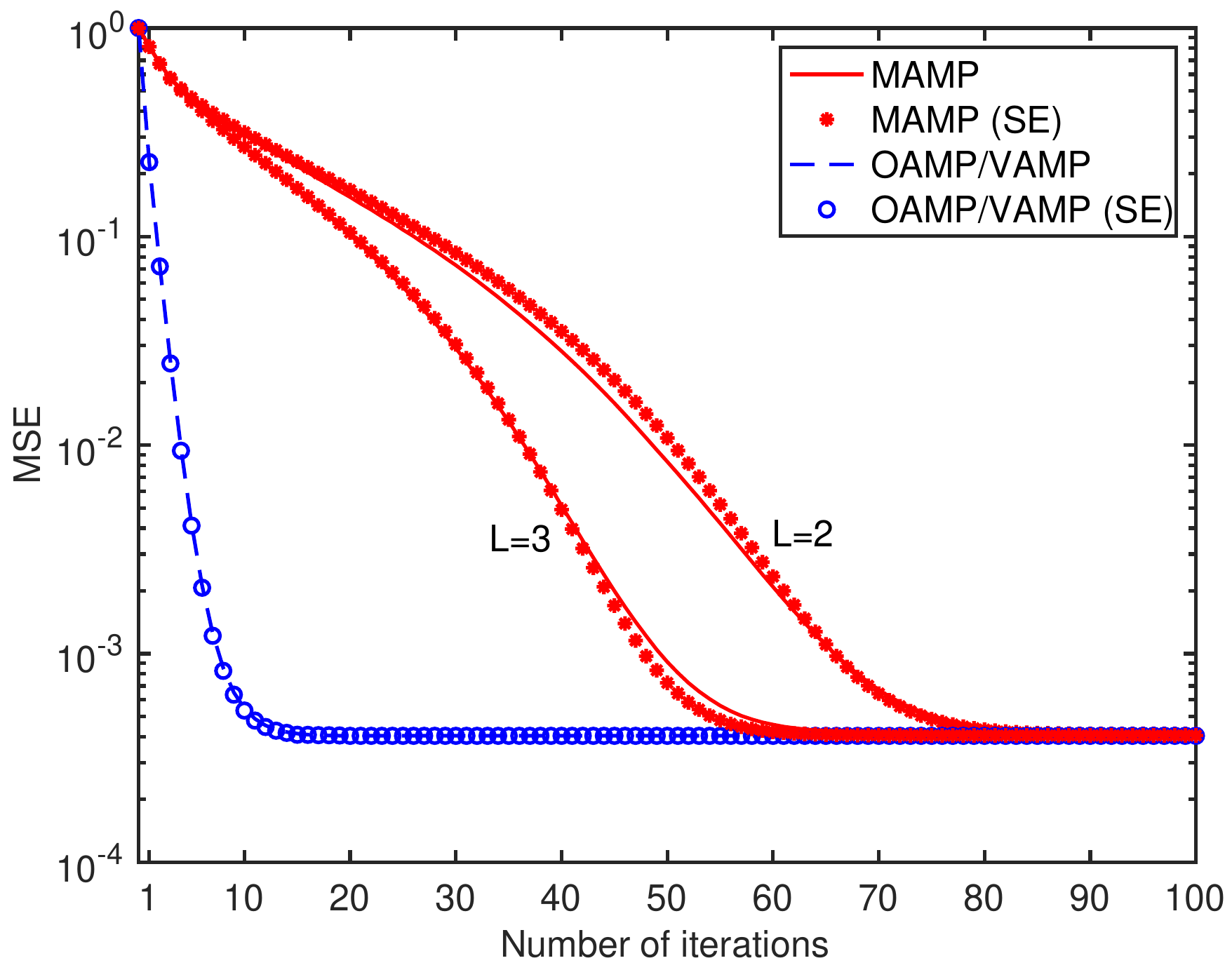} 
\caption{MSE versus the number of iterations for MAMP and OAMP/VAMP with different damping lengths. $M=4096, N=8192$, $\mu=0.1$, ${\rm SNR}=30$ dB, $\kappa=100$ and $L=2$ and 3.}
\label{Fig:MAMP_kappa}
\end{figure}

Fig.~\ref{Fig:MAMP_kappa}  shows MSE versus the number of iterations for MAMP with different damping lengths. As can be seen, MAMP converges to OAMP/VAMP for the matrix with high condition number $\kappa=100$, and the state evolution (SE) of MAMP matches well with the simulated MSE. Note that CAMP diverges when $\kappa>15$ (see Fig. 4 in \cite{Takeuchi2020CAMP}). In addition,  MAMP with $L=3$ (damping length) has significant improvement in convergence speed compared with $L=2$ when the condition number is large. It should be mentioned that $L=3$ is generally enough for MAMP, since the MSEs of MAMP are almost the same when $L\ge 3$. %Thus, we did not show the MSE curves of MAMP with $L\ge 4$. %This is an interesting observation, but we still have no mathematical explanation for this phenomenon. 
   
\section{Conclusions} 
This paper proposes a low-cost MAMP for high-dimensional linear systems with unitarily transform matrices. The proposed MAMP is not only Bayes-optimal, but also has comparable complexity to AMP. Specifically, the techniques of long memory and orthogonalization are used to achieve the Bayes-optimal solution of the problem with a low-complexity MF. The convergence of MAMP is optimized with some relaxation parameters and a damping vector. The optimized MAMP is guaranteed to converge to the high-complexity OAMP/VAMP for all right-unitarily-invariant matrices.

% \section*{Acknowledgment}  
% The authors would like to thank Keigo Takeuchi for kindly sharing us the source code and the simulation data of CAMP.

\end{document}